\documentclass[fleqn,twoside]{article}
\usepackage[headings]{espcrc2}

\readRCS
$Id: espcrc2.tex,v 1.2 2004/02/24 11:22:11 spepping Exp $
\usepackage{graphicx}
\usepackage[figuresright]{rotating}
\usepackage{epsfig}

\def\ra{\!\rightarrow\!}
\def\dbar{\overline{D}{}^{\,0}}

\def\dkkpp{$D^0\ra K^+K^-/\pi^+\pi^-$}
\def\cp{$CP$}
\def\cpv{$CPV$}

\def\hb{{\it HERA-B\/}}
\def\dkpi{$D^0\ra K^+\pi^-$}

\def\belle{Belle}
\def\babar{Babar}
\def\fbinv{~fb$^{-1}$}

\def\simge{\mathrel{%
   \rlap{\raise 0.511ex \hbox{$>$}}{\lower 0.511ex \hbox{$\sim$}}}}
\def\simle{\mathrel{
   \rlap{\raise 0.511ex \hbox{$<$}}{\lower 0.511ex \hbox{$\sim$}}}}

\newcommand{\AmS}{{\protect\the\textfont2
  A\kern-.1667em\lower.5ex\hbox{M}\kern-.125emS}}

\begin{document}

\runtitle{A Future High Statistics Charm Mixing Experiment 
Using the Fermilab Tevatron}

\title{
\vspace*{-0.40in} 
\mbox{\large \hskip5.18in UCHEP-08-08} \\
\vspace*{0.20in} 
A Future High Statistics Charm Mixing Experiment 
Using the Fermilab Tevatron}

\author{A. J. Schwartz\address{Physics Department, 
    University of Cincinnati, Cincinnati, Ohio 45221}}
       
\begin{abstract}
We present an idea for a future $D^0$-$\dbar$ mixing and \cp\ violation
experiment to run at the Fermilab Tevatron. We estimate that in three 
years of running, such an experiment could reconstruct an order of 
magnitude more flavor-tagged \dkpi\ decays than will be reconstructed 
by the $B$-factory experiments with their full data sets. The resulting
sensitivity to \cp-violating parameters $|q/p|$ and ${\rm Arg}(q/p)$
is calculated from a global fit to \cp-violating observables, and 
it is found to be much greater than current world sensitivity.
\vspace{1pc}
\end{abstract}

\maketitle

\section{INTRODUCTION}

We present an idea to use the Fermilab Tevatron
to produce very large samples of $D^*$ mesons that decay via 
$D^{*+}\ra D^0\pi^+,\ D^0\ra K^+\pi^-$~\cite{charge-conjugate}. 
The decay time distribution of the ``wrong-sign'' $D^0\ra K^+\pi^-$ 
decay is sensitive to $D^0$-$\dbar$ mixing parameters $x$ and~$y$. 
Additionally, comparing the $D^0$ decay time distribution to that 
for $\dbar$ allows one to measure or constrain the \cp-violating 
(\cpv) parameters $|q/p|$ and ${\rm Arg}(q/p)\equiv\phi$. This 
method has been used previously by Fermilab experiments  
E791~\cite{e791_kpi} and E831~\cite{e831_kpi} to search 
for $D^0$-$\dbar$ mixing. However, those 
experiments ran in the 1990's and reconstructed only a few hundred 
flavor-tagged $D^0\ra K^+\pi^-$ decays. Technological advances in 
vertexing detectors and electronics made since E791 and E831 ran 
now make a much improved fixed-target experiment possible. Here we 
estimate the expected sensitivity of such an experiment and compare 
it to that of the $B$ factory experiments \belle\ and \babar.
Those experiments have reconstructed several thousand signal decays 
and, using these samples along with those for \dkkpp, have made the 
first observation of $D^0$-$\dbar$ mixing~\cite{babar_kpi,belle_kk}. 
We also compare the estimated sensitivity to that of hadron
experiments CDF and LHCb.
Although we focus on measuring $x,y,|q/p|$, and $\phi$, a much 
broader charm physics program is possible at a Tevatron experiment.

\section{EXPECTED SIGNAL YIELD}

We estimate the signal yield expected by scaling from two previous
fixed-target experiments, E791 at Fermilab and \hb\ at DESY. These 
experiments had center-of-mass energies and detector geometries 
similar to those that a charm experiment at the Tevatron would have.

\subsection{Scaling from \emph{HERA-B}}

\hb\ took data with various trigger configurations. One configuration
used a minimum bias trigger, and from this data set the experiment
reconstructed $61.3\,\pm 13$ $D^*$-tagged ``right-sign''
$D^0\ra K^-\pi^+$ decays in $182\times 10^6$ hadronic interactions~\cite{herab_kpi}. 
This yield was obtained after all selection requirements were applied. 
Multiplying this rate by the ratio of doubly-Cabibbo-suppressed to 
Cabibbo-favored decays
$R^{}_D\equiv\Gamma(D^0\ra K^+\pi^-)/\Gamma(D^0\ra K^-\pi^+)=0.380\%$~\cite{dcstocf}
gives a rate of reconstructed, tagged \dkpi\ decays per hadronic interaction
of $1.3\times 10^{-9}$. To estimate the sample 
size a Tevatron experiment would reconstruct, we assume the experiment 
could achieve a similar fractional rate. If the experiment ran at an 
interaction rate of 7~MHz (which was achieved by \hb\ using a two-track 
trigger configuration), and took data for $1.4\times 10^7$ live seconds 
per year, then it would nominally reconstruct 
$(7{\rm\ MHz})(1.4\times 10^7)(1.3\times 10^{-9})(0.5)=64000$ flavor-tagged 
\dkpi\ decays per year, or 192000 decays in three years of running. 
Here we have assumed a trigger efficiency of 50\% relative to that 
of~\hb, as the trigger would need to be more restrictive than 
the minimum bias configuration of \hb.

\subsection{Scaling from E791}

Fermilab E791 was a charm hadroproduction experiment that took data
during the 1991-1992 fixed target run. The experiment ran with a modest 
transverse-energy threshold trigger, and it reconstructed 35 $D^*$-tagged
\dkpi\ decays in $5\times 10^{10}$ hadronic interactions~\cite{e791_kpi}.
This corresponds to a rate of $7\times 10^{-10}$ reconstructed decays per 
hadronic interaction. Assuming a future Tevatron experiment achieves
this fractional rate, one estimates a signal yield of
$(7{\rm\ MHz})(1.4\times 10^7)(7\times 10^{-10})=69000$ per year, 
or 207000 in three years. This value is similar to that obtained by scaling 
from \hb. We have assumed the same trigger\,+\,reconstruction efficiency 
as that of E791, for lack of better knowledge. We note that
E791 had an inactive region in the middle of the tracking stations 
where the $\pi^-$ beam passed through, and a future Tevatron 
experiment could avoid this acceptance loss. We do not include 
any improvement for this in our projection.

\section{{\boldmath COMPARISON WITH THE $B$ FACTORIES AND CDF}}

We compare these yields with those that will
be attained by the $B$ factory experiments after they have
analyzed all their data. The \belle\ experiment reconstructed 4024
$D^*$-tagged \dkpi\ decays in 400\fbinv\ of data~\cite{belle_kpi},
and it is expected to record a total of 1000\fbinv\ when it completes
running. This integrated luminosity corresponds to 10060 signal events. 

The \babar\ experiment reconstructed 4030 tagged \dkpi\ decays 
in 384\fbinv\ of data~\cite{babar_kpi}, and the experiment recorded 
a total of 484\fbinv\ when it completed running in early~2008. 
Thus the total \babar\ data set corresponds to 5080 signal 
events. Adding this to the estimated final yield from \belle\ 
gives a total of 15100 \dkpi\ decays. This is less than 8\% of 
the yield estimated for a Tevatron experiment in three years 
of running.

The CDF experiment has reconstructed 12\,700 tagged \dkpi\ decays 
in 1.5\fbinv\ of data~\cite{cdf_kpi}, and it is expected to record
a total of 7-8\fbinv\ when the Tevatron stops running. This data set 
would correspond to $\sim$\,64000 signal decays, which is similar 
to what a future fixed-target Tevatron experiment would record in 
one year of running. Such a sample from CDF would demonstrate the 
charm physics capability of a hadroproduction experiment at the
Tevatron.

The KEK-B accelerator where \belle\ runs is scheduled to be 
upgraded to a ``Super-$B$'' factory running at a luminosity 
of $\sim\!8\times 10^{35}$~cm$^{-2}$\,s$^{-1}$~\cite{superbelle}. There is
also a proposal to construct a Super-$B$ factory in Italy near 
the I.N.F.N. Frascati laboratory~\cite{superB}. An experiment at 
either of these facilities would reconstruct very large 
samples of $D^{*+}\ra D^0\pi^+,\,D^0\ra K^+\pi^-$ decays, and in fact 
the resulting sensitivity to $x'^2$ and $y'$ may be dominated 
by systematic uncertainties. This merits further study. We 
note that the systematic errors obtained at a future
Tevatron experiment are expected to be smaller than those
at an $e^+e^-$ collider experiment, due to the superior
vertex resolution and $\pi/K$ identification possible 
with a forward-geometry detector.

\section{COMPARISON WITH LHCb}

The LHCb experiment has a forward geometry
and is expected to reconstruct $D^{*+}\ra D^0\pi^+,\,D^0\ra K^+\pi^-$ 
decays in which the $D^*$ originates from a $B$ decay.
The resulting sensitivity to mixing 
parameters $x'^2$ and $y'$ has been studied in Ref.~\cite{lhcb_kpi}.
This study assumes a $b\bar{b}$ cross section of 500~$\mu$b and
estimates several unknown trigger and reconstruction efficiencies.
It concludes that approximately 58000 signal decays would be reconstructed
in 2\fbinv\ of data, which corresponds to one year of running. This 
yield is similar to that estimated for a Tevatron experiment. 
However, LHCb's trigger is efficient only for $D$ mesons having
high $p^{}_T$, i.e., those produced from $B$ decays. This
introduces two complications:
\begin{enumerate}
\item some fraction of prompt $\dbar\ra K^+\pi^-$ decays will be 
mis-reconstructed or undergo multiple scattering and, after being 
paired with a random soft pion, will end up in 
the \dkpi\ sample (fitted for $x'^2$ and $y'$). As the 
production rate of prompt $D$'s is $~$two orders of magnitude
larger than that of $B$'s, this component may be non-negligible
and thus would need to be well-understood when fitting.
\item to obtain the $D^*$ vertex position (i.e., the
origin point of the $D^0$), the experiment must reconstruct
a $B\ra D^* X$ vertex, and the efficiency for this is not 
known. Monte Carlo studies indicate it is 51\%~\cite{lhcb_kpi}, 
but there is uncertainty in this value.
\end{enumerate}

The LHCb study found that, for $N^{}_{K^+\pi^-}=232500$,
a signal-to-background ratio ($S/B$) of 0.40, and a decay
time resolution ($\sigma^{}_t$) of 75~ps, the statistical 
errors obtained for $x'^2$ and $y'$ were 
$6.4\times 10^{-5}$ and $0.87\times 10^{-3}$, respectively. 
These values are less than half of those that we estimate 
can be attained by the $B$ factories by scaling current 
errors by $\sqrt{N_{K^+\pi^-}}$:
$\delta x'^2\approx 14\times 10^{-5}$ and 
$\delta y'\approx 2.2\times 10^{-3}$.
As the signal yield, $S/B$, and $\sigma^{}_t$ of a future Tevatron 
experiment are similar to those for LHCb, we expect that similar 
errors for $x'^2$ and $y'$ can be attained.

\begin{figure}[htb]
\vbox{
\epsfig{file=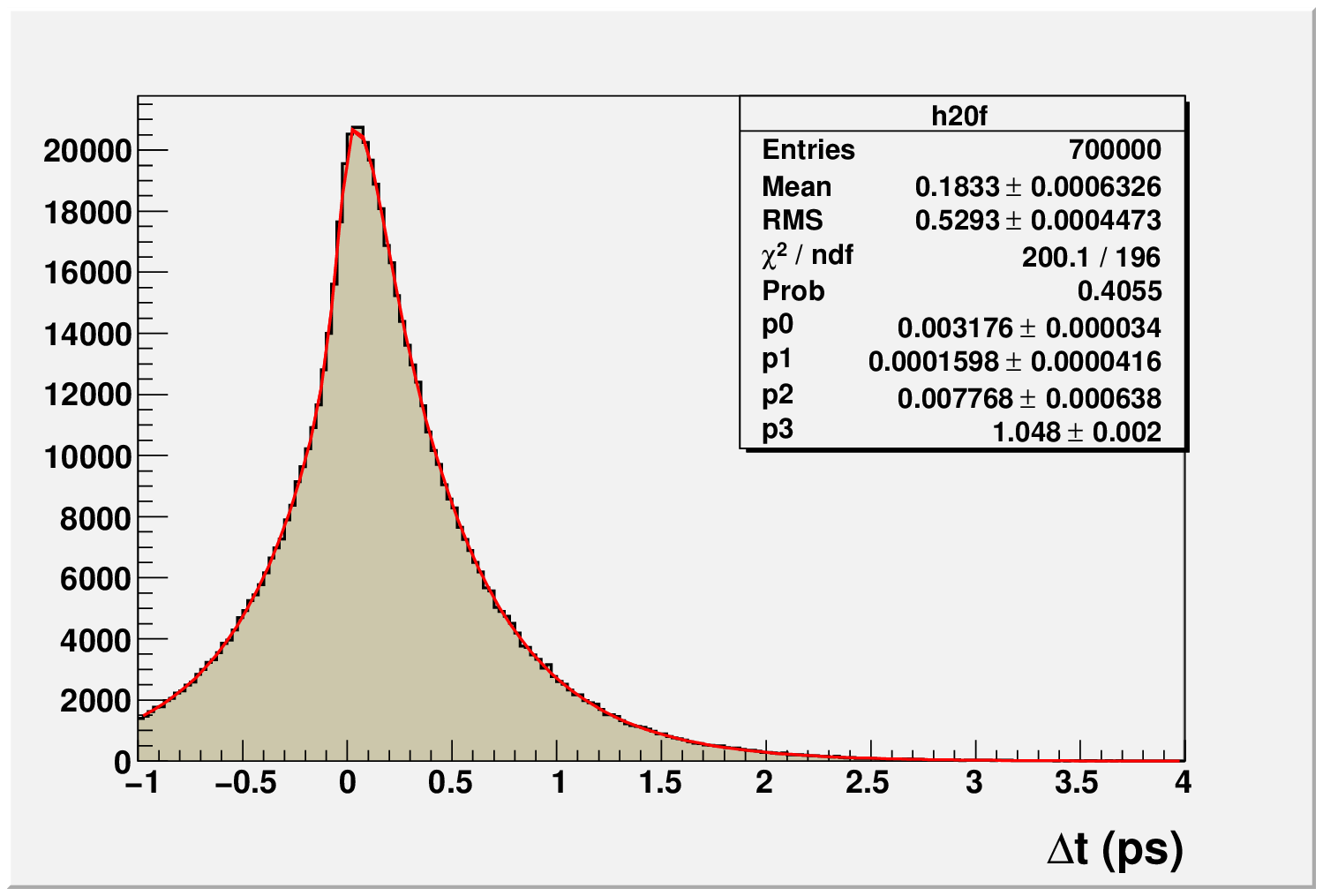,width=2.9in}
\vskip0.10in
\epsfig{file=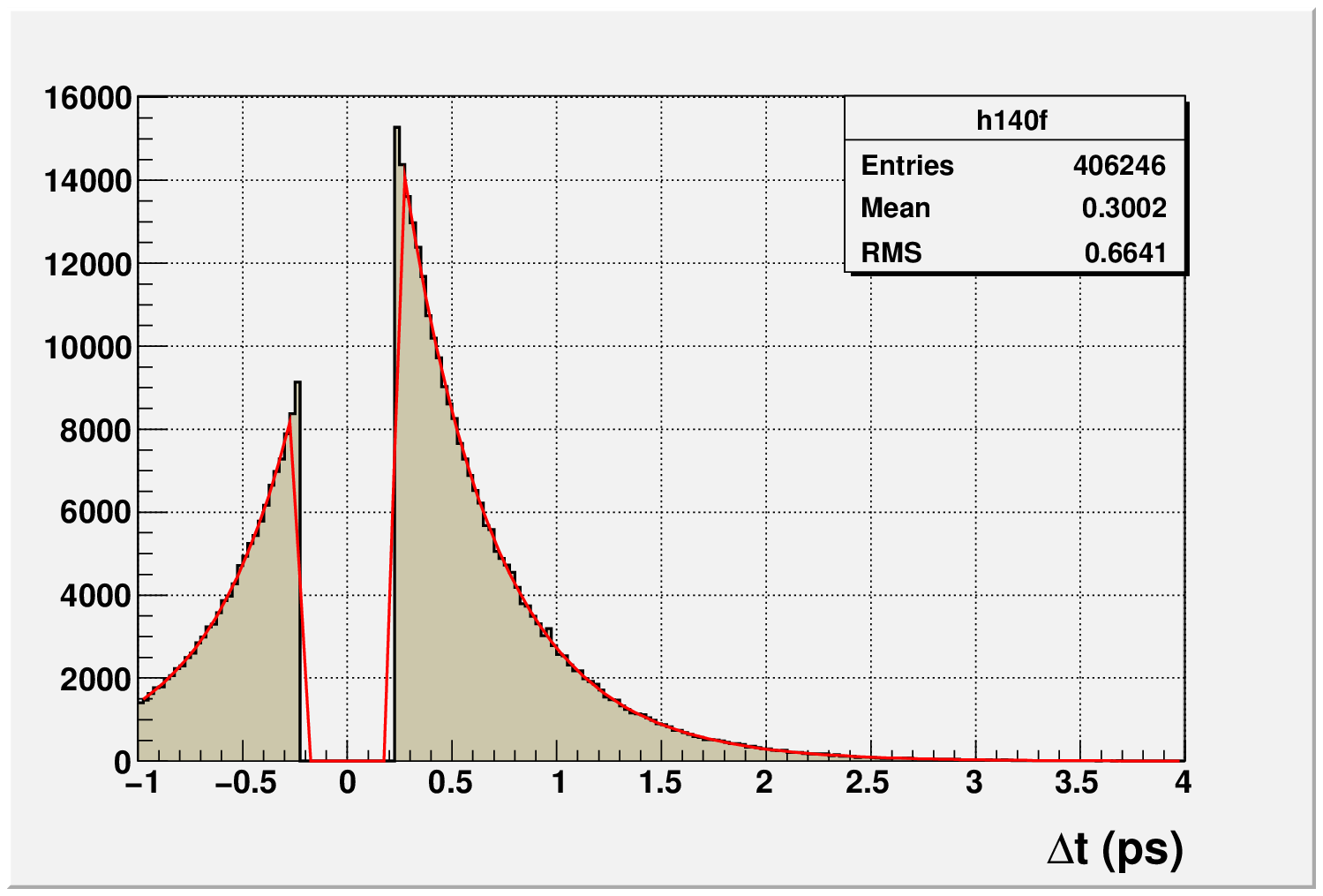,width=2.9in}
}
\vskip-0.30in
\caption{Monte Carlo \dkpi\ decay time distributions  
without (top) and with (bottom) a minimum decay time cut.
Superimposed is the result of a fit. The ratio of signal to
background after the $t^{}_{\rm min}$ ($=\!\tau^{}_D/2$) cut is 
0.40, and the decay time resolution $\sigma^{}_t$ is 75~fs.}
\label{fig:toymc_fit}
\end{figure}

To check these estimates, we have done a ``toy'' Monte Carlo (MC) 
study to estimate the sensitivity of a Tevatron experiment.
The results obtained are similar to those of LHCb:
for $N^{}_{K^+\pi^-}=200000$, $S/B=0.40$, $\sigma^{}_t=75$~ps,
and a minimum decay time cut of $0.5\times\tau^{}_D$
(to reduce combinatorial background), we find 
$\delta x'^2=5.8\times 10^{-5}$ and 
$\delta y'=1.0\times 10^{-3}$.
These errors are the RMS's of the distributions of
residuals obtained from fitting an ensemble of 200 experiments.
A typical fit is shown in Fig.~\ref{fig:toymc_fit}.

\section{{\boldmath GLOBAL FIT FOR \cpv\ PARAMETERS}}

If we assume the $\delta x'^2$ and $\delta y'$ errors
obtained in our toy MC study (which are close to the values
obtained in the LHCb study), we can estimate the resulting 
sensitivity to \cpv\ parameters $|q/p|$ and~$\phi$.
The first parameter characterizes \cpv\ in the mixing
of $D^0$ and $\dbar$ mesons, while the second parameter
is a phase that characterizes \cpv\ resulting from interference
between an amplitude with mixing and a direct decay amplitude.
In the Standard Model, $|q/p|$ and $\phi$ are essentially 
1 and 0, respectively; a measurable deviation from these 
values would indicate new physics.

To calculate the sensitivity to $|q/p|$ and $\phi$,
we do a global fit of eight underlying parameters to 28 
measured observables. The fitted parameters are $x$ and $y$,
strong phases $\delta^{}_{K\pi}$ and $\delta^{}_{K\pi\pi}$, $R^{}_D$,
and \cpv\ parameters $A^{}_D,\,|q/p|$ and~$\phi$.
Our fit is analogous to that done by the Heavy Flavor
Averaging Group (HFAG)~\cite{hfag_charm_fits}; the only
difference is that we reduce the errors for $x'^2$ and $y'$
according to our toy MC study, and we also reduce the error 
for $y^{}_{CP}$ by a similar fraction. This latter parameter is 
measured by fitting the decay time distribution of \dkkpp\ 
decays, which would also be triggered on and reconstructed 
by a Tevatron charm experiment.

\begin{figure}[htb]
\vbox{
\epsfig{file=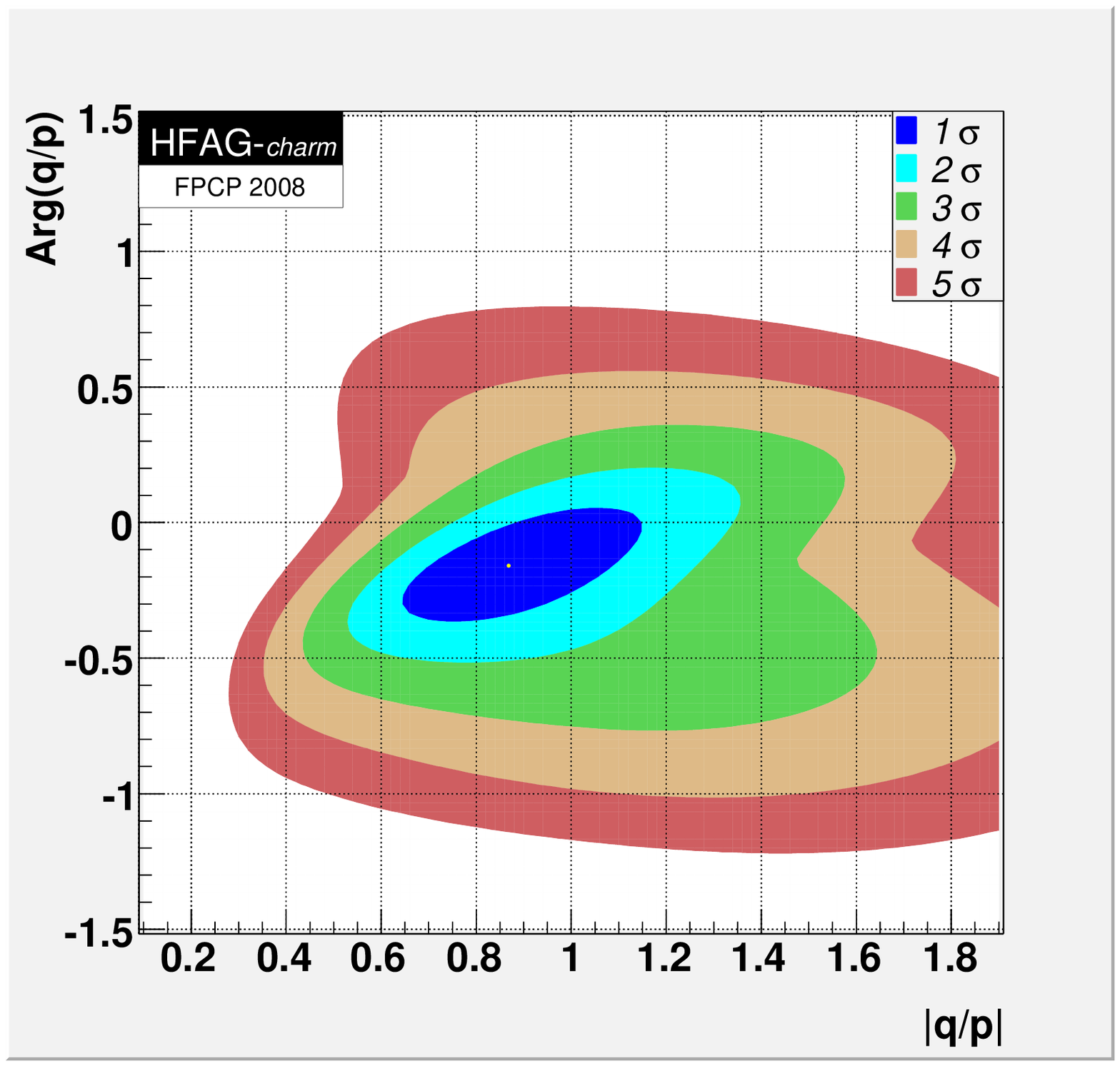,width=2.9in}
\vskip0.10in
\epsfig{file=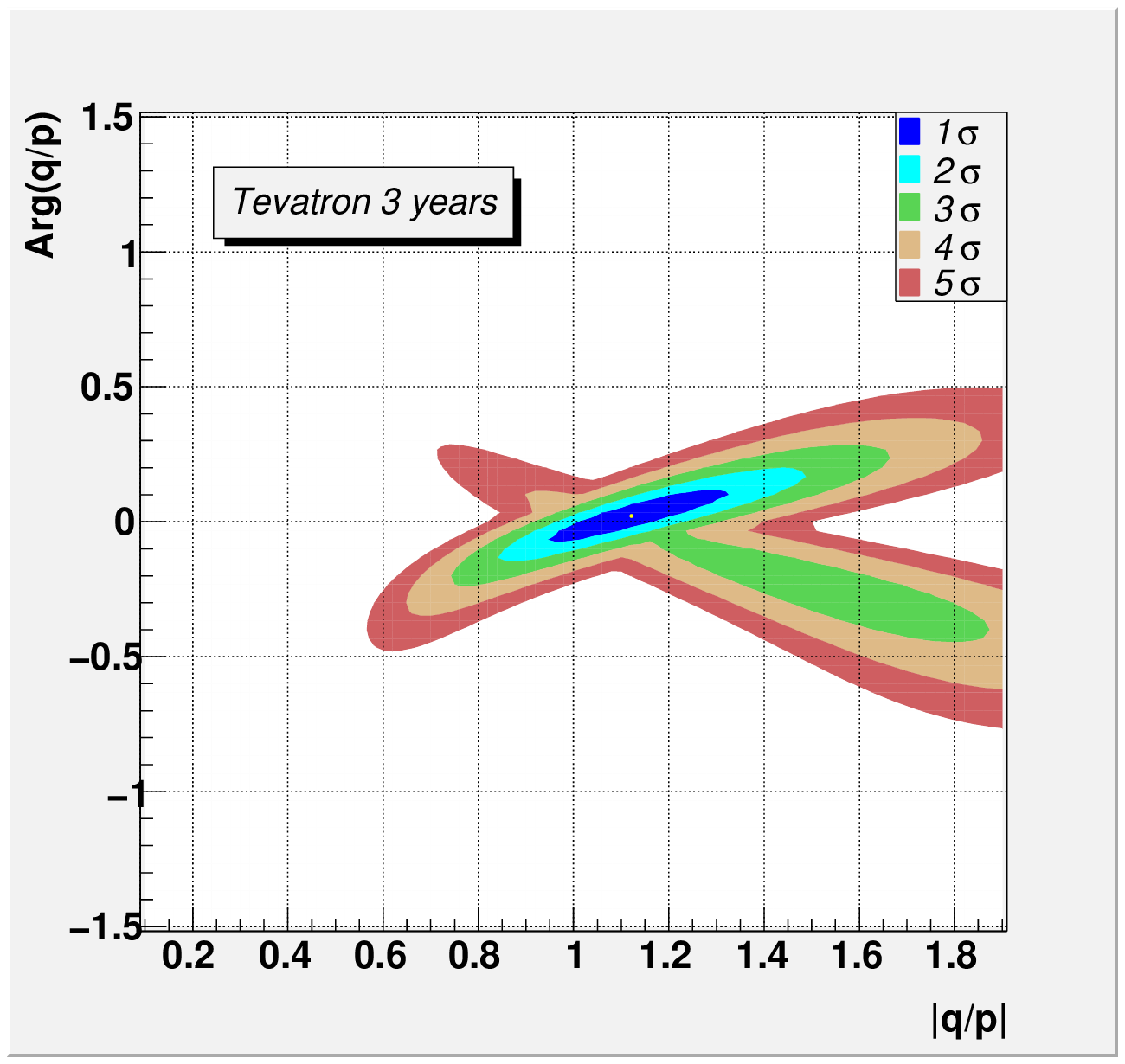,width=2.9in}
}
\vskip-0.30in
\caption{$|q/p|$ versus $\phi$ likelihood contours resulting 
from a global fit to measured observables (see text).
Top: data after FPCP 2008, from the Heavy Flavor Averaging
Group~\cite{hfag_charm_fits}. Bottom: after three years of 
running of a Tevatron charm experiment. }
\label{fig:fit_results}
\end{figure}

The results of the fit are plotted in Fig.~\ref{fig:fit_results}b. 
The figure shows two-dimensional likelihood contours for $|q/p|$ 
and~$\phi$; for comparison, the analogous HFAG plot is shown in 
Fig.~\ref{fig:fit_results}a. One sees that a future Tevatron 
experiment would yield a very substantial improvement.

\section{SUMMARY}

In summary, we note the following and conclude:
\begin{itemize}
\item $D^0$-$\dbar$ mixing is now established, and attention
has turned to the question of whether there is \cpv\ in this system. 
\item Technical advances in detectors and electronics made
since the last Fermilab fixed-target experiments ran would make
a new experiment much more sensitive to mixing and \cpv\ effects.
Silicon strips and pixels for vertexing are well-developed, and 
detached-vertex-based trigger concepts and prototypes exist 
(e.g., \hb, CDF, BTeV, LHCb).
\item Such an experiment would have substantially better sensitivity
to mixing and \cpv\ than all \belle\ and \babar\ data together 
will provide. The Tevatron data should have less background than 
LHCb data. Systematic uncertainties may also be less than those 
of the $B$ factory experiments and LHCb.
\item The Tevatron and requisite beamlines are essentially available.
\item Such an experiment could help untangle whatever signals for new 
physics appear at the Tevatron or LHC.
\end{itemize}
Recently, a working group has formed to study the physics potential
of a charm experiment at the Tevatron in more detail. Information
about this working group and its results can be obtained at
{\tt http://www.nevis.columbia.edu/twiki/bin/ view/FutureTev/WebHome}.



\begin{thebibliography}{99}
\bibitem{charge-conjugate}
Charge-conjugate modes are implicitly included unless noted otherwise.
\end{thebibliography}
\end{document}